\documentclass[draftcls,onecolumn,12pt,journal]{IEEEtran}

\usepackage{cite}
\usepackage[cmex10]{amsmath}

\usepackage[caption=false,font=footnotesize]{subfig}

\usepackage{fixltx2e}

\usepackage{amsfonts,amsthm}
\usepackage{graphicx,color}
\usepackage{comment}
\usepackage{enumerate}
\newcommand{\abs}[1]{|#1|}
\newcommand{\norm}[1]{\|#1\|}
\newcommand{\bas}[1]{\begin{align*}#1\end{align*}}
\newcommand{\ba}[1]{\begin{align}#1\end{align}}
\newcommand{\R}{\mathbb{R}}

\newcommand{\N}{\mathcal{N}}

\newcommand{\fee}{\varphi}

\renewcommand{\sf}{s_{ \mathrm{firm } }}
\newcommand{\Sf}{S_{ \mathrm{firm } }}
\newcommand{\gf}{g_{ \mathrm{firm } }}
\newcommand{\Gf}{G_{ \mathrm{firm } }}
\newcommand{\F}{\mathcal{F}}

\theoremstyle{plain}
\newtheorem{theorem}{Theorem}[section]
\newtheorem{lemma}[theorem]{Lemma}
\newtheorem{proposition}[theorem]{Proposition}
\newtheorem{corollary}[theorem]{Corollary}

\theoremstyle{definition}
\newtheorem{definition}[theorem]{Definition}

\theoremstyle{remark}
\newtheorem{remark}[theorem]{Remark}

\DeclareMathOperator*{\argmin}{arg\,min}
\DeclareMathOperator{\sign}{sign}
\DeclareMathOperator{\prox}{prox}
\DeclareMathOperator{\supp}{supp}

\begin{document}
\title{Compressed Sensing Recovery\\ via Nonconvex Shrinkage Penalties}

\author{Joseph~Woodworth, Rick~Chartrand~\IEEEmembership{Senior~Member,~IEEE}% <-this % stops a space
\IEEEcompsocitemizethanks{\IEEEcompsocthanksitem J. Woodworth is with the Department
of Mathematics, University of California, Los Angeles,
CA, 90095, USA; %\protect\\
e-mail: jwoodworth@math.ucla.edu
\IEEEcompsocthanksitem R. Chartrand is with  the Theoretical Division, MS B284, Los Alamos National Laboratory, Los Alamos, NM, 87545, USA; e-mail: rickc@lanl.gov}% <-this % stops a space
\thanks{Manuscript received ????, 2014; revised ????.}}


\markboth{IEEE Transactions on Information Theory}%
{Woodworth \MakeLowercase{\textit{et al.}}: Compressed Sensing Recovery Via Nonconvex Penalties}

%\IEEEpubid{\makebox[\columnwidth]{\hfill 0000--0000/00/\$00.00~\copyright~2012 IEEE}%


\IEEEtitleabstractindextext{%
\begin{abstract}
%No more that 200 words, passive voice
% The below is ~112 words, i.e. I could roughly double it
The $\ell^0$ minimization of compressed sensing is often relaxed to $\ell^1$, which yields easy computation using the shrinkage mapping known as soft thresholding, and can be shown to recover the original solution under certain hypotheses.  Recent work has derived a general class of shrinkages and associated nonconvex penalties that better approximate the original $\ell^0$ penalty and empirically can recover the original solution from fewer measurements.  We specifically examine \emph{$p$-shrinkage} and \emph{firm thresholding}.  In this work, we prove that given data and a measurement matrix from a broad class of matrices, one can choose parameters for these classes of shrinkages to guarantee exact recovery of the sparsest solution.  We further prove convergence of the algorithm \emph{iterative $p$-shrinkage} (IPS) for solving one such relaxed problem.
\end{abstract}

}

\maketitle

% is not selected <OR> if conference mode is selected - because compsoc
% conference papers position the abstract like regular (non-compsoc)
\IEEEdisplaynontitleabstractindextext

% \fi
\IEEEpeerreviewmaketitle

\section{Introduction}
% Computer Society journal papers do something a tad strange with the very
% first section heading (almost always called "Introduction"). They place it
% ABOVE the main text! IEEEtran.cls currently does not do this for you.
% However, You can achieve this effect by making LaTeX jump through some
% hoops via something like:
%
%\ifCLASSOPTIONcompsoc
%  \noindent\raisebox{2\baselineskip}[0pt][0pt]%
%  {\parbox{\columnwidth}{\section{Introduction}\label{sec:introduction}%
%  \global\everypar=\everypar}}%
%  \vspace{-1\baselineskip}\vspace{-\parskip}\par
%\else
%  \section{Introduction}\label{sec:introduction}\par
%\fi
%
% Admittedly, this is a hack and may well be fragile, but seems to do the
% trick for me. Note the need to keep any \label that may be used right
% after \section in the above as the hack puts \section within a raised box.


\IEEEPARstart{C}{ompressed} sensing has been successfully applied in a multitude of scientific fields, ranging from image processing tasks to radar to coding theory, making the potential impact of advancements in theory and practice rather large.  Compressed sensing methods rely on the notion of sparsity, which is primarily approximated via the $\ell^1$ norm \cite{candes-2006-robust,donoho-2006-compressed}.  The nature and limitations of this relaxation have been well-studied \cite{bruckstein2009sparse,cai2013sharp,davies2009restricted,donoho2003optimally,elad2002generalized,foucart2010note,foucart2012sparse,gribonval2003sparse,gribonval-2007-highly}, as well as some alternative relaxations, such as the $\ell^p$ quasinorm \cite{chartrand-2008-restricted,wu2013improved,aldroubi2011stability,chen2010lower,davies2009restricted,foucart2009sparsest,gribonval2003sparse,gribonval-2007-highly,lai2011new,saab-2008-stable,sun2011sparse,sun2012recovery}.  The nonconvex $\ell^p$ quasinorm approaches present a tradeoff: closer approximation of sparsity for harder analysis and computation.  Recent work has introduced generalized nonconvex penalties \cite{chartrand-2009-fast,chartrand-2013-generalized,chartrand-2014-shrinkage,chartrand-2013-nonconvex,chartrand-2013-nonconvex2,chartrand-2012-nonconvex,antoniadis-2007-wavelet} that have thus far demonstrated strong empirical performance \cite{chartrand-2009-fast,voronin-2013-new,chartrand-2014-shrinkage,chartrand-2013-nonconvex2}.  In this paper, we prove conditions that guarantee good performance of these generalized penalties.


\subsection{Compressed Sensing}

%\IEEEpubidadjcol

Compressed sensing seeks to represent a signal from a small number of linear measurements.  
We let the vector $x\in\R^n$ represent the original signal.  
The linear measurements are the result of an application of the short and fat measurement matrix $A\in\R^{m\times n}$, with $m\ll n$.  
One is given the measurements $b:=Ax$ and wants to recover $x$.
Of course $m\ll n$ implies that $Ax=b$ is an underdetermined linear system in $x$, so additional assumptions must be made about $x$.
Thus one assumes that $x$ is \emph{sparse}, meaning that it has few nonzero entries.
By considering the standard definition of $p$ norms for vectors,
\ba{
\|w\|_p^p
:=
\sum_i \abs{ w_i }^p,
}
and taking the limit as $p$ approaches 0 from above, we get the $\ell^0$ penalty, $\|w\|_0$, which counts the number of nonzero entries of $w$.  % \comm{Need $p$th power for limit to be correct.}
One would like to find the sparsest vector $w\in\R^n$ whose measurements are $b$, which suggests the following optimization problem:
\ba{\label{eq:0}
\min_w \|w\|_0 \text{ subject to }Aw=b.
}

Unfortunately, this problem is known to be NP-hard (Non-deterministic Polynomial-time hard) in general \cite[Sec. 9.2.2]{muthukrishnan2005data}.  In other words, without making further assumptions on $A$ and $x$, an algorithm solving this problem would be computationally intractable.  For this reason, one relaxes the problem, replacing the $\ell^0$ penalty with other penalties.

\subsection{$\ell^1$ relaxation}

The $\ell^1$ relaxed version of the compressed sensing problem is as follows:
\ba{\label{eq:1}
\min_w \|w\|_1 \text{ subject to }Aw=b.
}
% \textbf{Rick, can you help me say the right things here?  E.g. It's not plainly an LP, but what is the complexity?}
In contrast to the combinatorial $\ell^0$ problem, this problem minimizes a convex energy subject to linear constraints, and can be recast as a linear program. % \comm{Split w into positive and negative parts to get linear program, see GPSR paper.}
Extensive theory has been developed to study the properties of solutions to convex problems \cite{boyd2009convex}.  Further, a subproblem related to the $\ell^1$ relaxation of compressed sensing has a closed-form solution, given by an application of a shrinkage operator:
\begin{definition}
\label{def:softthresh}
Soft thresholding is given by the following formula:
\begin{equation}
  S_{ \lambda, 1 }( x )_i = s_{\lambda,1}( \abs{ x_i } ) \sign( x_i )=\max \{ \abs{ x_i } - \lambda, 0 \} \sign( x_i ).
\end{equation}
\end{definition}
The role soft thresholding plays is as the \emph{proximal mapping} of the $\ell^1$ norm:
\begin{equation}\label{eq:prox1}
  S_{ \lambda, 1 }( x ) = \prox_\lambda \norm{ \cdot }_1 ( x ) := \argmin_w \lambda \norm{ w }_1 + \tfrac{ 1 }{ 2 } \norm{ w - x }_2^2.
\end{equation}
Several algorithms for compressed sensing make use of this proximal mapping, such as iterative soft thresholding \cite{daubechies-2004-iterative}, alternating direction method of multipliers (ADMM) \cite{glowinski-1975-approximation,gabay-1976-dual,goldstein-2009-split,boyd2011distributed}, and the Chambolle-Pock algorithm \cite{chambolle2011first}. The explicit formula for \eqref{eq:prox1} makes the use of $\ell^1$ regularization particularly convenient.

All of this suggests why the $\ell^1$ relaxation of compressed sensing is nice to solve, but does not motivate it as the right problem to solve.  In particular, one is interested in conditions under which the solution to the $\ell^1$ relaxation \eqref{eq:1} of compressed sensing equals or approximately equals the solution of the original $\ell^0$ compressed sensing problem \eqref{eq:0}. The papers \cite{candes-2006-robust,donoho-2006-compressed} developed theory for the recovery of the $\ell^0$ solution by the $\ell^1$ problem.  In the years the followed, getting looser conditions for exact $\ell^1$ recovery received continuing interest \cite{bruckstein2009sparse,cai2013sharp,davies2009restricted,donoho2003optimally,elad2002generalized,foucart2010note,foucart2012sparse,foucart2009sparsest,gribonval2003sparse,gribonval-2007-highly}.   One type of condition for recovery of the $\ell^0$ solution from the $\ell^1$ problem relies on the \emph{restricted isometry constants} associated with the measurement matrix $A$.  The restricted isometry constant of order $k$ associated with the matrix $A\in\R^{m\times n}$ is the smallest $\delta_k\ge 0$ such that the following holds for all $x\in\R^{n}$ with $\|x\|_0\le k$ \cite{CandesTao1}:
\ba{
(1-\delta_k)\|x\|_2^2
\le
\|Ax\|_2^2
\le
(1+\delta_k)\|x\|_2^2.
}
Note that when $\delta_k> 1$ the lower bound becomes trivial and the upper bound can be improved by rescaling $A$.  Thus any measurement matrix, with appropriate rescaling, can achieve $\delta_{k}=1$, so one typically only regards $\delta_k\in [0,1).$  One of the best current $\ell^1$ recovery results states that for sufficiently large $n$, a sparse vector $x\in \R^n$ with $\|x\|_0=k$ can be recovered by $\ell^1$ minimization as long as $k<m/2$ and the restricted isometry constant of order $2k$ associated with $A$ satisfies $\delta_{2k}\le 1/2$ \cite{cai2013sharp}.

% http://www.math.tu-berlin.de/fileadmin/i26_fg-kutyniok/Kutyniok/Papers/SurveyCS.pdf
% http://arxiv.org/pdf/math/0409186.pdf

\subsection{$\ell^p$ relaxations ($0<p<1$)}

A similar relaxation of the $\ell^0$ problem that achieves recovery results in broader cases is $\ell^p$ minimization for $0<p<1$.  In contrast to the $\ell^1$ norm, the $\ell^p$ quasinorms for $0<p<1$ are not convex.  Hence much of the theory of convex analysis no longer applies, making solution uniqueness and convergence results more complicated.  However, the loss of convexity comes with the benefit that $\ell^p$ is better able to approximate the original $\ell^0$ than $\ell^1$ can.  As a result, one can show that for any given measurement matrix with restricted isometry constant $\delta_{2k}<1$, there exists some $p\in(0,1)$ that will guarantee exact recovery of signals with support smaller than $k<m/2$ by the $\ell^p$ minimization problem \cite{wu2013improved}. It has also been demonstrated empirically that $\ell^p$ minimization gives better sparse recovery results than $\ell^1$ minimization \cite{chartrand-2007-exact,chartrand-2008-nonconvex,chartrand-2007-nonconvex}, with improved robustness \cite{saab-2008-stable,aldroubi2011stability,sun2011sparse}.
% \textbf{What to say about the cusp? (as opposed to the bounded derivatives of generalized shrinkage penalties), make it clear why this is not the end of the story}

Consider the proximal mapping of the $\ell^p$ quasinorm (to the $p^{ \text{th} }$ power, for simplicity), that is, 
\begin{equation}\label{eq:proxp}
  \prox_\lambda \norm{ \cdot }_p^p ( x ) := \argmin_w \lambda \norm{ w }_p^p + \tfrac12 \norm{ w - x }_2^2.
\end{equation}
Unfortunately, \eqref{eq:proxp} is a discontinuous mapping \cite{yukawa-2013-lp}, and there is no closed-form expression for \eqref{eq:proxp} for general $p$. (The expression given in \cite{majumdar-2011-algorithm} is incorrect. For the special cases of $p = 1 / 2$ or $2 / 3$, the proximal mapping can be expressed in terms of the solution of a cubic or quartic equation, explicitly but cumbersomely.) This prevents several efficient algorithms from being generalized from $\ell^1$ to $\ell^p$ minimization.

\subsection{Generalized shrinkage}

The need for an explicit proximal mapping motivates the approach of specifying a shrinkage mapping, and minimizing an implicitly-defined penalty function whose proximal mapping is the specified shrinkage \cite{antoniadis-2007-wavelet,chartrand-2009-fast,chartrand-2013-generalized,chartrand-2014-shrinkage}. In this work, we extend theoretical results for recovery of sparse signals to the case of penalty functions induced by two families of shrinkages, $p$-shrinkage and firm thresholding (see Defs.~\ref{def:pshrink},~\ref{def:firm} below). In Section~\ref{sec:nonconvpen} we describe these shrinkage mappings, and how they are the proximal mappings of nonconvex penalty functions. In Section~\ref{sec:ERP} we prove conditions for the exact recovery of sparse signals via minimizing such nonconvex penalty functions. In Section~\ref{sec:stability} we demonstrate the stability of signal recovery to noisy measurements and approximately sparse signals, and in Section~\ref{sec:convergence} we show the algorithmic convergence of \emph{iterative $p$-shrinkage} (IPS).

% Note that IEEE typically puts floats only at the top, even when this
% results in a large percentage of a column being occupied by floats.
% However, the Computer Society has been known to put floats at the bottom.

% An example of a double column floating figure using two subfigures.
% (The subfig.sty package must be loaded for this to work.)
% The subfigure \label commands are set within each subfloat command,
% and the \label for the overall figure must come after \caption.
% \hfil is used as a separator to get equal spacing.
% Watch out that the combined width of all the subfigures on a 
% line do not exceed the text width or a line break will occur.
%
%\begin{figure*}[!t]
%\centering
%\subfloat[Case I]{\includegraphics[width=2.5in]{box}%
%\hfil
%\subfloat[Case II]{\includegraphics[width=2.5in]{box}%
%\label{fig_second_case}}
%\caption{Simulation results.}
%\label{fig_sim}
%\end{figure*}
%
% Note that often IEEE papers with subfigures do not employ subfigure
% captions (using the optional argument to \subfloat[]), but instead will
% reference/describe all of them (a), (b), etc., within the main caption.

\section{Generalized shrinkage penalties}
\label{sec:nonconvpen}

As described above, nonconvex penalty functions have been shown both theoretically and empirically to give better results for compressed sensing than the $\ell^1$ norm. In order to make use of any of several efficient algorithms, we wish to consider penalty functions with explicit proximal mappings. In this section, we consider two such families of functions.

\subsection{$p$-shrinkage and firm thresholding}

\begin{figure}[t]
  \centering
  \includegraphics[width=0.5\textwidth]{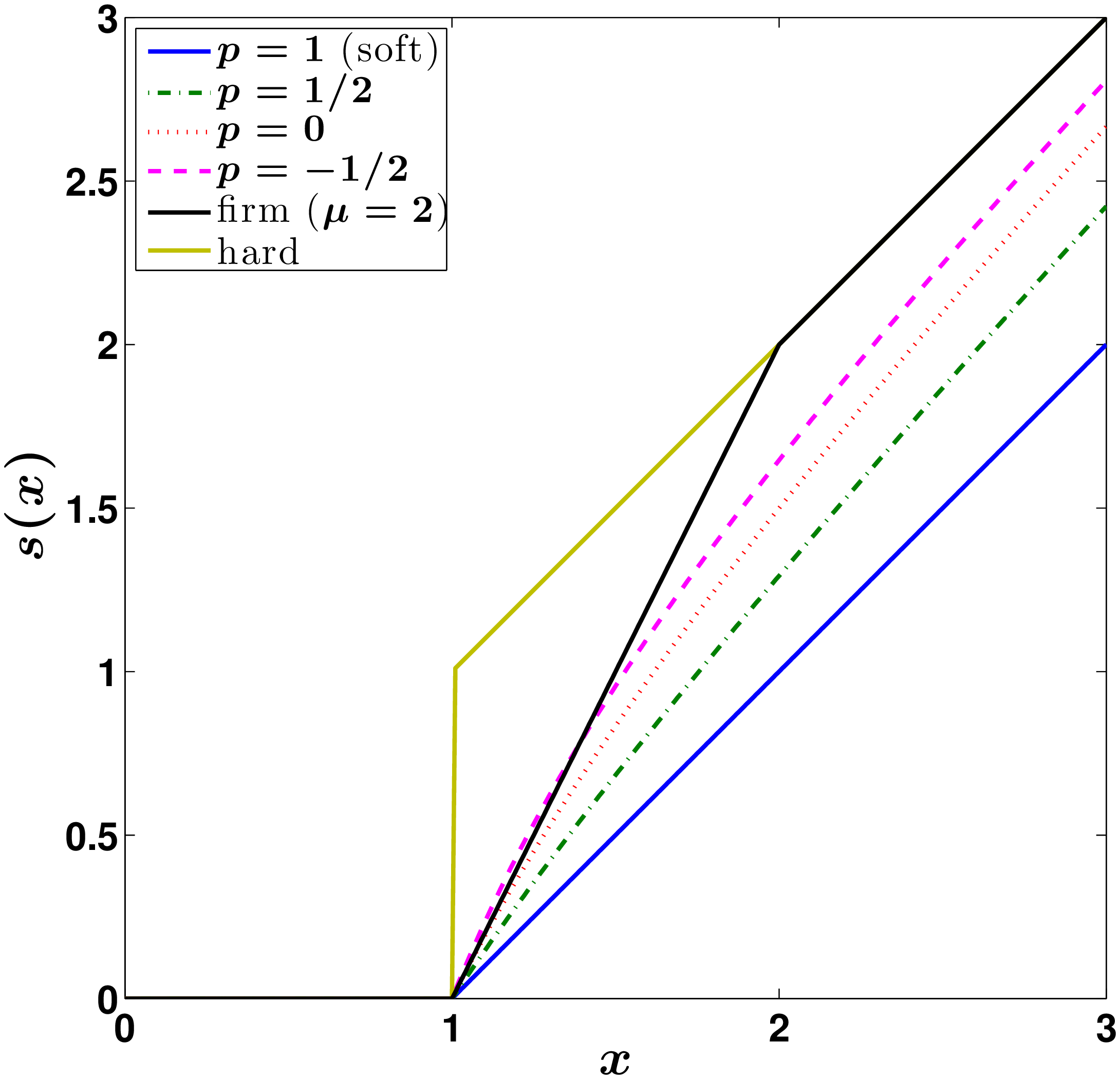}
  \caption{Plot of several shrinkage functions, all with $\lambda = 1$. The smaller the value of $p$, the smaller the bias applied to large inputs. Firm thresholding removes the bias completely for large enough inputs, without the discontinuity of hard thresholding.}\label{fig:shrinks}
\end{figure}

First we consider a shrinkage mapping, a version of which first appeared in \cite{chartrand-2009-fast}, that has some qualitative resemblance to the $\ell^p$ proximal mapping, while being continuous and explicit:
\begin{definition}
\label{def:pshrink}
For $\lambda > 0$, the $p$-shrinkage mapping $S_p = S_{ \lambda, p }$ for $p \in \R$ is defined by $S_p(x)_i = s_p ( \abs{ x_i } ) \sign( x_i )$, where the shrinkage function $s_p = s_{ \lambda, p }$ is defined by
\ba{\label{eq:pshrink}
s_p ( t ) 
=
\max
\{
t - \lambda^{2-p} t^{p-1},0
\}.
}
\end{definition}
See Fig.~\ref{fig:shrinks} for example plots. When $p = 1$, $p$-shrinkage and soft thresholding coincide. The smaller the value of $p$, the less $p$-shrinkage shrinks large inputs. In the limit as $p \rightarrow -\infty$, $p$-shrinkage tends pointwise to \emph{hard thresholding}:
\begin{definition}
\label{def:firm}
For $\lambda > 0$, the hard thresholding mapping $H_\lambda$ is defined by
\ba{\label{eq:hard}
H_{\lambda}(x)_i
=
\begin{cases}
  0 & \text{ if } \abs{ x_i } \le \lambda, \\
  x_i & \text{ if } \abs{ x_i } > \lambda.
\end{cases}
}
\end{definition}
Hard thresholding is related to the proximal mapping of the $\ell^0$ penalty function:
\begin{equation}\label{eq:prox0}
  H_{ \sqrt{ 2 \lambda } } \in \prox_\lambda \norm{ \cdot }_0,
\end{equation}
the right side of \eqref{eq:prox0} being two-valued in components satisfying $x_i^2 = 2 \lambda$. Hard thresholding imposes no bias on large inputs, but its discontinuity makes it very unstable when used with ADMM \cite{dong-2013-efficient}.

Another shrinkage mapping we consider is \emph{firm thresholding}, a continuous, piecewise-linear approximation of hard thresholding. Firm thresholding was first introduced in \cite{gao-1997-waveshrink} in connection with the WaveShrink procedure for denoising and non-parametric regression. It was not known at the time to be the proximal operator of a given penalty function.
\begin{definition}
  For $\lambda > 0$ and $\mu > \lambda$, the firm thresholding mapping $\Sf = S_{ \lambda, \mu, \mathrm{firm} }$ is defined by $\Sf ( x )_i = \sf( \abs{ x_i } )$, where $\sf = s_{ \lambda, \mu, \mathrm{firm} }$ is defined by
  \ba{\label{eq:firm}
    \sf ( t ) 
    =
    \begin{cases}
      0 & \text{ if } t \le \lambda, \\
      \frac{ \mu }{ \mu - \lambda } ( t - \lambda ) & \text{ if } \lambda \le t \le \mu, \\
      t & \text{ if } t \ge \mu.
    \end{cases}
  }
\end{definition}
Note that $S_{ \lambda, \lambda, \mathrm{firm} } = H_\lambda$, and $\lim_{ \mu \rightarrow \infty } S_{ \lambda, \mu, \mathrm{firm} }( x ) = S_{ \lambda, 1 }( x )$ pointwise. Thus both $p$-shrinkage and firm thresholding can be seen as generalizing both soft and hard thresholding.

\subsection{Shrinkage-induced penalty functions}

\begin{figure}[t]
  \centering
  \includegraphics[width=0.7\textwidth]{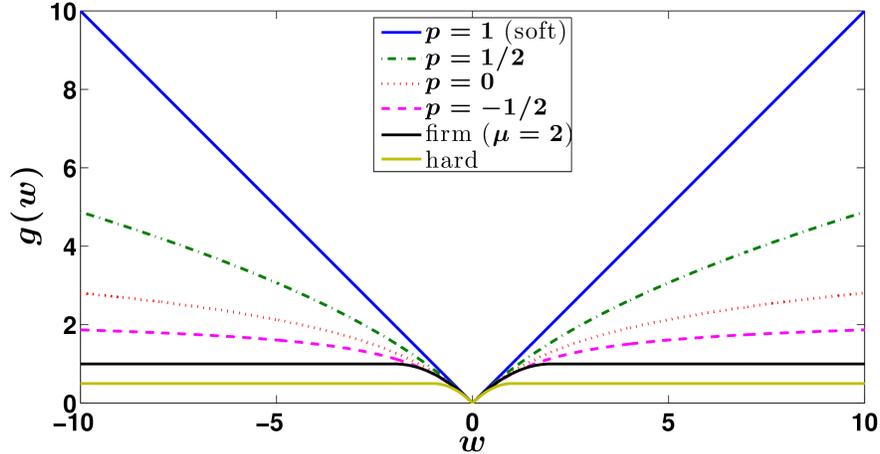}
  \caption{Plot of penalty component function $g$ induced by several shrinkage mappings, all with $\lambda = 1$. The smaller the value of $p$, the slower the growth of the $p$-shrinkage penalty function, being bounded above when $p < 0$. Both firm and hard thresholding have penalty functions that are quadratic near the origin, then constant.}\label{fig:penalties}
\end{figure}

Our motivation for considering alternative shrinkage mappings is to have them as closed-form proximal mappings. This requires that the shrinkages actually be the proximal mappings of penalty functions. The following theorem guarantees this. It is proved in \cite[Thm.~1]{chartrand-2014-shrinkage}, and strengthens the earlier result of Antoniadis \cite[Prop.~3.2]{antoniadis-2007-wavelet}.
\begin{theorem}
\label{thm:genshrink}
Suppose $s:[0,\infty)\rightarrow \R$ is continuous, satisfies $x\le \lambda\Rightarrow s(x)=0$ for some $\lambda\ge 0$, is strictly increasing on $[\lambda,\infty),$ and $s(x)\le x.$  Define $S(x)_i=s(|x_i|)\sign(x_i),$ for each $i$.  Then $S$ is the proximal mapping of a penalty function $G(w)=\sum_i g(w_i)$ where $g$ is even, strictly increasing and continuous on $[0,\infty),$ differentiable on $(0,\infty),$ and nondifferentiable at 0 iff $\lambda>0$ (in which case $\partial g( 0 ) = [ -1, 1 ]$).  If also $x-s(x)$ is nonincreasing on $[\lambda,\infty),$ then $g$ is concave on $[0,\infty)$ and $G$ satisfies the triangle inequality.
\end{theorem}
Both $p$-shrinkage and firm thresholding satisfy all hypotheses of the theorem for all parameter values. The proof of the theorem constructs $g$ using the Legendre-Fenchel transform \cite{rockafellar-1998-variational} of an antiderivative of $s$.
Because of the nature of the Legendre-Fenchel transform, this often does not produce a closed-form expression for $g$. We consider this as an acceptable price to pay for having an explicit proximal mapping, which is much more useful for most of today's state-of-the-art algorithms for compressed sensing than having an explicit penalty function. In the case of the penalty function $G_p$ induced by $p$-shrinkage, we can compute $g_p( w )$ numerically, and example plots are in Fig.~\ref{fig:penalties}. In addition to the properties guaranteed by Thm.~\ref{thm:genshrink}, it can be shown that $\lim_{ w \rightarrow \infty } g_p( w ) - w^p / p - C_p = 0$ for $p \ne 0$ and constant $C_p$ depending only on $p$. This includes $p < 0$, in which case it follows that $g_p( w )$ is bounded above. For $p = 0$, we have $\lim_{ w \rightarrow \infty } g_0( w ) - \log w - C = 0$ instead.

In the case of the penalty function $\Gf$ induced by firm thresholding, $\gf$ does have a closed form:
\begin{equation}\label{eq:gfirm}
  \gf( w ) = 
  \begin{cases} 
    \abs{ w } - w^2 / ( 2 \mu ) & \text{ if } \abs{ w } \le \mu, \\
    \mu / 2 & \text{ if } \abs{ w } \ge \mu.
  \end{cases}
\end{equation}
Note that $\gf( w )$ is independent of $\lambda$, except that $\mu \ge \lambda$ is required by the definition of $\gf$.

Although the statement of Thm.~\ref{thm:genshrink} excludes hard thresholding (being discontinuous), the construction in the proof does produce a penalty function $G_{ \mathrm{hard} }$. It coincides with $\Gf$ for $\mu = \lambda$. The part of the conclusion of the theorem that doesn't hold is that $\prox_\lambda G_{ \mathrm{hard} }( \lambda )$ is the entire interval $[ 0, \lambda ]$, while $H_\lambda( \lambda )$ is generally defined to take on a single value from this interval (namely 0 in our definition \eqref{eq:hard}).

\subsection{Example}

To motivate the consideration of $p$-shrinkage and firm thresholding, we consider a generalization of an example appearing in the first compressed sensing paper \cite{candes-2006-robust}. We seek to reconstruct the $256 \times 256$ Shepp-Logan phantom image from samples of its 2-D discrete Fourier transform (DFT), taken along radial lines, thereby simulating both MRI and X-ray CT data (the latter by way of the Fourier slice theorem). See Fig.~\ref{fig:SL}. Since the phantom has a sparse gradient, we seek to solve the following optimization problem:
\begin{equation}\label{eq:gexample}
  \min_x G( \nabla x ) \text{ subject to } \F x = b,
\end{equation}
where $G$ is one of the penalty functions being compared, $\nabla$ is a discrete gradient using forward differences and periodic boundary conditions, $\F$ is the 2-D DFT, and $b$ contains the sample data. We solve \eqref{eq:gexample} with ADMM, where the shrinkage mapping is $p$-shrinkage with $p \le 1$ or firm thresholding. See \cite{chartrand-2013-nonconvex2} for details, being also a straightforward generalization of the algorithm of \cite{goldstein-2009-split}.

With $G = G_1 = \norm{ \cdot }_1$, 18 lines are required for exact reconstruction, while using $G = G_{ -1 / 2 }$, 9 lines suffice, as shown in \cite{chartrand-2009-fast}, the latter being the fewest that had been demonstrated at that time. In \cite{chartrand-2013-generalized} (see also \cite{chartrand-2014-shrinkage}), 6 lines were shown to suffice using the $G$ induced by a shrinkage mapping that is a $C^\infty$ approximation of hard thresholding. This is the fewest possible, since with 5 lines, there are fewer measurements than nonzero gradient pixels, so that the phantom will not even be a local minimizer of the problem with $G = \norm{ \cdot }_0$. However, here we report that using $G = \Gf$ (with $\lambda = 0.1$ and $\mu = 2.5$), 6 lines also suffice, and many fewer ADMM iterations are needed (337 versus 2213).

\newlength{\figsl}
\setlength{\figsl}{0.235\textwidth}
\begin{figure}[t]
  \centering
  \subfloat[Shepp-Logan phantom]{\includegraphics[width=\figsl]{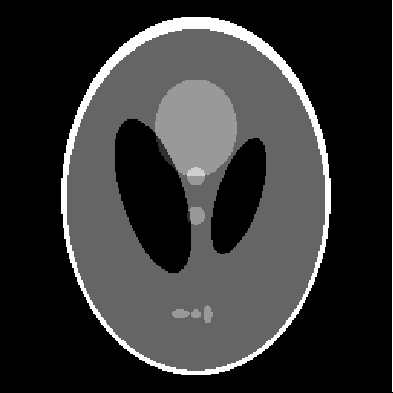}}\;
  \subfloat[$p = 1$, 18 lines]{\includegraphics[width=\figsl]{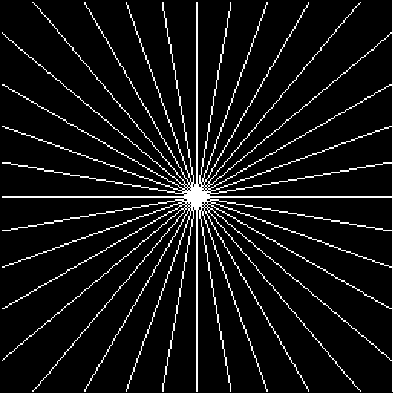}}\;
  \subfloat[$p = -1 / 2$, 9 lines]{\includegraphics[width=\figsl]{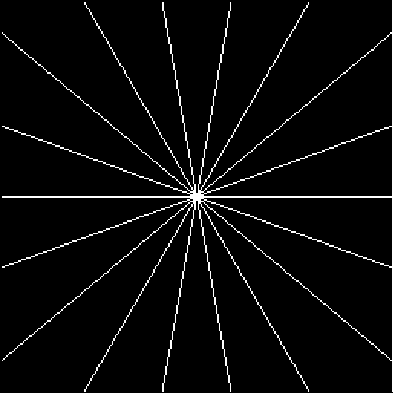}}\;
  \subfloat[firm, 6 lines]{\includegraphics[width=\figsl]{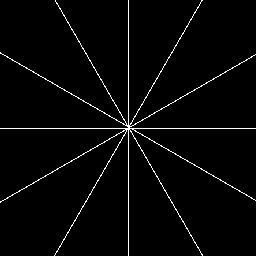}}
  \caption{The Shepp-Logan phantom, and the number of radial lines of Fourier samples needed to reconstruct the phantom perfectly using different penalty functions.}\label{fig:SL}
\end{figure}

While this example is an ideal case, using a very sparse image and noisefree measurements, this does demonstrate that $p$-shrinkage and firm thresholding induce penalty functions that can be useful for recovering sparse signals. Now we turn to a theoretical analysis of the sparse recovery performance of minimizing these penalty functions.

\section{Exact recovery}
\label{sec:ERP}
In this section, we establish sufficient conditions for exact recovery of sparse signals from noisefree measurements by solving a minimization problem with penalty function $G$:
\begin{equation}\label{eq:geq}
  \min_w G( w ) \text{ subject to } A w = b.
\end{equation}
Our objective is to determine sufficient conditions in the case where $G$ is a penalty function induced by a shrinkage mapping; however, we will establish conditions for a somewhat more general class of penalty functions $G$. We shall assume that the measurement matrix $A\in \R^{m\times n}$ has the Unique Representation Property (URP), i.e., any $m$ columns of $A$ are linearly independent.  This implies that any vector in $\ker(A)$ has at least $m + 1$ nonzero entries. The URP can be regarded as a \emph{generic} property of matrices; for example, a matrix whose entries are independently and identically distributed samples drawn from any absolutely continuous probability distribution will have URP with probability 1.

\begin{remark}
The URP implies that the $m$ rows of $A$ are linearly independent.  
Thus an orthonormal basis for the span of the rows can be formulated as linear combinations of the rows of $A$.
So if we multiply $A$ by a product of elementary matrices, $E$, corresponding to the necessary elementary row operations, the resulting product will have orthonormal rows.  
Since elementary matrices are invertible, $Aw=b$ is equivalent to $EAw=Eb$.  
Also, since each elementary matrix is invertible, $A_T$  being full rank for $|T|=m$ implies  $EA_T$ is full rank as well, and so $A$ satisfying the URP implies  $EA$ satisfies the URP.
Thus we can always transform the problem so that the rows of $A$ are orthonormal, i.e.,  $AA^T=I$, and so without loss of generality, we assume that the $A$ given satisfies $AA^T=I$.
\end{remark}

We shall also assume that $G(w)=\sum_i g(w_i)$ with 
\begin{enumerate}[I)]
\item $g(0)=0$, and $g$ even on $\R$; and
\item $g$ is continuous on $\R$, and either strictly increasing and strictly concave on $\R$, or strictly increasing and strictly concave on $( 0, \gamma ]$ and constant on $[ \gamma, \infty )$ for some $\gamma > 0$.
\end{enumerate}
These conditions imply that $g$ is nondecreasing and concave on $[ 0, \infty )$, is everywhere nonnegative, and satisfies the triangle inequality.

\begin{lemma}\label{lem:fp}
  The penalty functions $\Gf$ and $G_p$ (for $-\infty<p<1$) satisfy the above conditions.
\end{lemma}
\begin{IEEEproof}
  It is clear from the expression \eqref{eq:gfirm} for $\gf$ that $\Gf$ satisfies the conditions with $\gamma=\mu$.

  For $G_p$, by Thm.~\ref{thm:genshrink} we get condition I, and that $g_p$ is differentiable on $( 0, \infty )$ with $g_p' > 0$. It suffices to prove that $g_p$ is twice differentiable on $( 0, \infty )$ with $g_p'' < 0$; it will be no more difficult to show that $g_p \in C^\infty ( 0, \infty )$. We need some details from the construction of $g_p$, from \cite{chartrand-2014-shrinkage}. We have 
\begin{equation}\label{eq:gp}
  g_p( w ) = ( f_p^*( w ) - w^2 / 2 ) / \lambda,
\end{equation}
  where $f_p' = s_p$ and $f_p^*$ is the Legendre-Fenchel transform of $f_p$. Since $s_p$ is continuous and nondecreasing, $f_p$ is $C^1$ and convex. Then by \cite[Prop.~11.3]{rockafellar-1998-variational}, we have that 
  \begin{equation}\label{eq:dual}
    x \in \partial f_p^*( w ) \Leftrightarrow w = f_p'( x ) = s_p( x ).
  \end{equation}

  Fix $w > 0$, and let $x$ be such that $w = s_p(x)$. From \eqref{eq:pshrink}, we must have $x > \lambda$, so $w = x - \lambda^{ 2 - p } x^{ p - 1 }$. If we define $F( x, w ) = x - \lambda^{ 2 - p } x^{ p - 1 } - w$, we have that $F( \cdot, w )$ is $C^\infty$ on $( 0, \infty )$, and $\frac{ \partial^k F }{ \partial x^k }( x, w ) \ne 0$ for $x \in ( \lambda, \infty )$. Thus by the implicit function theorem, $f_p^*$ is $C^\infty$ on $( 0, \infty )$, hence $g_p$ is as well by \eqref{eq:gp}.

  Returning to $w = x - \lambda^{ 2 - p } x^{ p - 1 }$, by \eqref{eq:gp}, \eqref{eq:dual}, and the differentiability of $f_p^*$, we have
  \ba{
    g_p'(w)
    =
    ( ( f_p^* )'( w ) - w ) / \lambda 
    =
    ( \lambda / x )^{1-p}.
  }
  Thus $g_p'(w)$ is decreasing in $x$ on $(\lambda,\infty)$, and since $x$ is a strictly increasing function of $w$ on $(0,\infty)$, $g_p''(w)<0$ on $(0,\infty)$.
\end{IEEEproof}

%% We are interested in showing conditions under which 
%% \bas{
%% x^*
%% =\argmin
%% G(w)
%% \text{ s.t. }Aw=b
%% }
%% recover the $k$-sparse, sparsest solution $x$ such that $Ax=b$.

\begin{lemma}
\label{mnzs}
  Assume $A\in\R^{m\times n}$ satisfies the URP and $G$ satisfies (I,II) above. Then the global minimizer of \eqref{eq:geq} has $m$ or fewer nonzero entries.
\end{lemma}
\begin{IEEEproof}
  Consider $w$ such that $Aw=b$ and $\|w\|_0 > m$. Define the matrix $M$ to have the columns $-w_ie_i$. The set of vectors $Mv$ with $\supp(v)\subset \supp(w)$ span a subspace of dimension greater than $m$. Since $\dim( \ker(A) )= n - m$, we can choose a $v$ with $Mv\in \ker(A)$ and $\|v\|_\infty= 1$.  

  For all $t \in \R$, $w + t M v$ is feasible. Define $T = \{ i : v_i \ne 0 \text{ and } \abs{ w_i } < \gamma \}$ (taking $\gamma = +\infty$ if the first case of assumption II holds). First suppose $T \ne \O$. Then by assumption II, the function $t \mapsto G( w + t M v )$ is strictly concave on an interval $[ -\delta, \delta ]$, with $\delta > 0$ chosen small enough that every $( w + t M v )_i$ has the same sign as $w_i$ for all $\abs{ t } \le \delta$. Then $G( w ) > \min \{ G( w - \delta M v ), G( w + \delta M v ) \}$, and $w$ is not a global minimizer.

  Otherwise, we have $v_i \ne 0 \Rightarrow \abs{ w_i } \ge \gamma$. Let $t_0 = \sup \{ t : \forall i \min \{ \abs{ ( w - t M v )_i },  \abs{ ( w + t M v )_i } \} \ge \gamma \}$. Then taking $t_1 = t_0 + \delta$ with $\delta > 0$ again small enough that every $( w \pm t_1 M v )_i$ has the same sign as $w_i$, then one of $\abs{ ( w \pm t_1 M v )_i }$ is less than $\gamma$ for at least one $i$, giving a smaller value of $g$. Since all other components keep $g$ constant, we have one of $G( w \pm t_1 M v )$ being smaller than $G( w )$.
\end{IEEEproof}

\begin{lemma}
\label{alphabeta}
Assume $A\in\R^{m\times n}$ satisfies the URP.
Then the magnitudes of nonzero entries of vectors $y$ satisfying $Aw=b$ with $m$ or fewer nonzero entries are uniformly bounded below by some positive constant $\alpha$ and uniformly above by some positive constant $\beta$.
\end{lemma}

\begin{IEEEproof}
By the URP, every $m$ columns of $A$ can admit no more than one solution.  Thus there are no more than ${ n \choose m }$ vectors $w$ satisfying $Aw=b$ with $m$ or fewer nonzero entries.  Thus the set of nonzero entries of these vectors is finite and bounded below and above by $\alpha,\beta$ respectively.  Neither constant depends on $G$ in any way.
  \end{IEEEproof}
Note that Lemma~\ref{mnzs} and Lemma~\ref{alphabeta}  imply that the global
minimizer of the equality-constrained $G$ minimization problem has
nonzero entries with magnitude bounded below by
$\alpha$ and above by $\beta$.  

Next we introduce the $G$ Nullspace Property, a generalization of the $\ell^1$ Nullspace Property introduced in \cite{cohen-2009-compressed} for norms and implicitly in \cite{gribonval-2007-highly} for penalty functions belonging to a particular class. We denote $\{ 1, 2, \ldots, n \} = [ n ]$, and $T^c$ denotes the complement of $T$ in $[ n ]$.
\begin{definition}
The \emph{$G$ Nullspace Property} (or \emph{$G$ NSP}) of order $k$ for the matrix $A$ is satisfied when for all $h\in \ker(A) \backslash \{0\}$ and $T\subset [n]$ with $|T|\le k$ , one has $G(h_T)<G(h_{T^c}).$
\end{definition}

\begin{proposition}
\label{GNSPGERP}
For a penalty function $G$ satisfying the triangle inequality, the $G$ NSP implies exact recovery.
\end{proposition}
\begin{IEEEproof}
  We simply observe that the proof of \cite{gribonval-2007-highly} works assuming only that the penalty function satisfies the triangle inequality.
%% For given $b$, let $x^*$ be the global minimizer of the equality-constrained $G$ minimization problem and $x$, the sparsest feasible vector with $\|x\|_0=k$.  
%% Suppose we do not have exact recovery of $x^*$, i.e $h=x^*-x\in ker(A) \backslash\{0\}$.  
%% We shall use the triangle inequality for $G$ and the fact that $G(x^*)\le G(x).$  
%% Define $T=supp(x)$.
%% \bas{
%% G(h_{T^C})
%% =
%% G(x^*_{T^C})
%% =&
%% G(x^*_{T^C})
%% +G(x^*_T-x^*_T+x)
%% -G(x)\\
%% %
%% \le&
%% G(x^*_{T^C})
%% +G(x^*_T)+G(x^*_T-x)
%% -G(x)\\
%% %
%% \le&
%% G(x^*)
%% -G(x)
%% +G(h_T)
%% \le G(h_T) 
%% }
%% Thus no exact recovery implies no $G$ NSP.  So $G$ NSP implies exact recovery.
\end{IEEEproof}

\begin{definition}
  Let the matrix $A\in \mathbb R^{m\times n}$  and the vector $b\in\mathbb R^m$  be given. Let $x$ be the sparsest solution to $Aw=b$, $k = \norm{ x }_0$ with $2 k \le m$, and $T=\supp(x)$. We say the \emph{G Restricted Nullspace Property} (or \emph{G RNSP}) of order $k$ is satisfied if whenever $w$ satisfies $Aw=b$ and $\|w\|_0 \le m$, then for $h = x - w$, we have either $h = 0$ or $G(h_T)<G(h_{T^c})$.
\end{definition}

Note that the $G$ NSP of order $k$ for $A$ implies the $G$ RNSP of order $k$ for $A$.  However, examining the proof of Proposition~\ref{GNSPGERP} from \cite{gribonval-2007-highly} and applying Lemma~\ref{mnzs} shows that in fact $G$ RNSP suffices for exact recovery. We assume $2 k \le m$ to guarantee that the sparsest solution of $A w = b$ is unique, as URP ensures that a second solution must have more than $m - k$ nonzero components.

\begin{proposition}
For penalty function $G$ satisfying the triangle inequality, $G$ RNSP implies exact recovery.
\end{proposition}

\begin{theorem}[$G$ exact recovery]
\label{GERP}
Assume $A\in\R^{m\times n}$ satisfies the URP and $G$ satisfies (I,II) above.  
For given $b$, let $x^*$ be the global minimizer of \eqref{eq:geq} and $x$ the sparsest feasible vector. Let $k = \|x\|_0$, and  
define $\alpha,\beta$ to be the lower and upper bound of magnitudes of nonzero entries of feasible vectors with $m$ or fewer nonzero components as in Lemma~\ref{alphabeta}.  If $2k \le m$ and $kg(2\beta)<(m+1-k)g(\alpha)$ then $x^*=x$.
\end{theorem}

\begin{IEEEproof}
  Let $h = x^* - x$. Since $x$ is supported on $T$, $h_{T^c}=x^*_{T^c}$, and so for all $t\in T^c$, $|h(t)|$ is either zero or at least $\alpha$.  Also, since $h\in \ker(A)$, if $h\ne 0$, then $\|h_{ T^c }\|_0 \ge m + 1 - k$ (otherwise we would have $\norm{ h }_0 \le m$, violating URP), so that $G(h_{T^c})\ge (m+1-k) g(\alpha)$. Also,
  \begin{equation}
    G( h_T ) \le \sum_{ i \in T } g( \abs{ x^*_i } + \abs{ x_i } ) \le k g( 2 \beta ) < ( m + 1 - k ) g( \alpha )
  \end{equation}
  by assumption. Thus either $h = 0$ or $G(h_T)<G(h_{T^c})$, so G RNSP is satisfied.
\end{IEEEproof}

\begin{corollary}[$\Gf$ exact recovery]
\label{GFERP}
  Assume $A\in\R^{m\times n}$ satisfies URP and $G=\Gf$, the penalty corresponding to firm thresholding.
  For given $b$, let $x^*$ be the global minimizer of \eqref{eq:geq} and $x$ the sparsest feasible vector. Let $k = \|x\|_0$.  
  If $2k \le m$ and 
  \begin{equation}
    \mu < \min\biggl\{ \alpha \frac{m+1-k}{k} \biggl( 1+\sqrt{1-\frac{k}{m+1-k}} \biggr) , 2\beta \biggr\},
  \end{equation} 
  then $x^*=x$.
\end{corollary}
\begin{IEEEproof}
  Since $A$ satisfies URP and $G$ satisfies (I,II), we may apply Theorem~\ref{GERP}.  The inequality conditions from Theorem~\ref{GERP} are $2k \le m$ and $kg(2\beta) < (m+1-k)g(\alpha)$.  We know $\alpha<2\beta$.  If we have $\mu \le \alpha$, then the inequality becomes $k\mu/2 < (m+1-k)\mu/2$ which follows automatically from $2k \le m$.  And so we satisfy the hypotheses of Theorem~\ref{GERP}, and thus have exact recovery.  If instead we have $\alpha<\mu<2\beta$, we can evaluate the desired inequality as follows:
\begin{gather}
  \frac{k\mu}{2} 
  \le
  (m+1-k)(\alpha-\alpha^2/2\mu),
  \\
  \mu^2
  -
  2\frac{m+1-k}{k}
  \alpha \mu
  +\frac{m+1-k}{k}
  \alpha^2
  < 0,\\
  \biggl|\mu -\alpha \frac{m+1-k}{k}\biggr|
  <
  \alpha\frac{m+1-k}{k}\sqrt{1-\frac{k}{m+1-k}},\\
  \alpha \frac{m+1-k}{k}\biggl(1-\sqrt{1-\frac{k}{m+1-k}}\biggr)
  <
  \mu
  <
  \alpha \frac{m+1-k}{k}\biggl(1+\sqrt{1-\frac{k}{m+1-k}}\biggr).
\end{gather}
The left bound is always looser than the assumed $\alpha <\mu$ (for $2k<m+1$), so the condition $\mu<\alpha \frac{m+1-k}{k}\biggl(1+\sqrt{1-\frac{k}{m+1-k}}\biggr)$ gives the desired inequality and guarantees exact recovery.
\end{IEEEproof}

\begin{corollary}[$G_p$ exact recovery]
\label{GPERP}
Assume $A\in\R^{m\times n}$ satisfies the URP and $G=G_p$, the $p-$shrinkage penalty.
For given $b$, let $x^*$ be the global minimizer of \eqref{eq:geq} and $x$ the sparsest feasible vector. Let $k = \|x\|_0$.  
If $2k \le m$ then there exist $\lambda>0$ and $0<p<1$ sufficiently small that $x^*=x$.  For any $p<0$ there also exists $\lambda>0$ sufficiently small that $x^*=x$.
\end{corollary}
\begin{IEEEproof}
Since $A$ satisfies the URP and $G_p$ satisfies (I,II), we may apply Theorem~\ref{GERP}.  The inequality conditions from Theorem~\ref{GERP} are $2k \le m$ and $kg(2\beta) < (m+1-k)g(\alpha)$.

Fix $w > 0$. As in the proof of Lemma~\ref{lem:fp}, we have 
\begin{equation}\label{eq:gpagain}
  g_p( w ) = ( f_p^*( w ) - w^2 / 2 ) / \lambda,
\end{equation}
where $f_p' = s_p$ and $f_p^*$ is the Legendre-Fenchel transform of $f_p$ and is smooth at $w$. Let $x = ( f_p^* )'( w )$, noting that while $w$ is fixed, $x$ depends on $\lambda$ and $p$. By \eqref{eq:dual}, we have $s_p(x)=w$, so that 
\begin{equation}\label{eq:xwsp}
  x - w = \lambda^{ 2 - p } x^{ p - 1 }.
\end{equation}
Furthermore, by \cite[Prop.~11.3]{rockafellar-1998-variational}, we have
\begin{equation}\label{eq:argdual}
  x = \argmin_x \bigl( x w - f_p( x ) \bigr),
\end{equation}
so that by definition of the Legendre-Fenchel transform,  
\begin{equation}\label{eq:argLF}
  f_p^*( w ) = x w - f_p( x ).
\end{equation}
Combining \eqref{eq:gpagain}, \eqref{eq:xwsp}, and \eqref{eq:argLF}, we obtain
\ba{
g_p( w ) 
&= 
( x w - f_p( x ) - w^2 / 2 ) / \lambda \nonumber \\
&= 
\bigl( x w - x^2 / 2 + \lambda^{ 2 - p } x^p / p - \lambda^2 ( 1 / p - 1 / 2 ) - w^2 / 2 \bigr) / \lambda \label{eq:gpcalc}\\
&=
 \lambda^{ 1 - p } x^p / p - ( x - w )^2 / ( 2 \lambda ) - \lambda ( 1 / p - 1 / 2 ) \nonumber \\
&=
\frac{\lambda}{p} \left(x/\lambda\right)^p  -\frac{\lambda}{2} (x/\lambda)^{2p-2}- \lambda ( 1 / p - 1 / 2 ). \label{eq:gpexpression}
}
(In \eqref{eq:gpcalc}, the expression for $f_p( x )$ is obtained by antidifferentiating $s_p$ with $f_p( 0 ) = 0$.)

\paragraph{Case $0 < p < 1$}
We want to show that for sufficiently small $0<\lambda$ and $0<p<1$, $g(2\beta)/g(\alpha)<(m+1-k)/k$.  By hypothesis, $(m+1-k)/k > 1$.  So it suffices to show for any fixed $\alpha, \beta$ with $0<\alpha<2\beta$, that $g(2\beta)/g(\alpha)\rightarrow 1$ as $(p,\lambda)\rightarrow (0^+,0^+).$  

By \eqref{eq:xwsp}, $x > w$ for any $\lambda$ and $p$, so $\lim_{\lambda\rightarrow 0^+}(x/\lambda)= \infty$. Then for $p<1$, 
\begin{equation}
  \lim_{ \lambda \rightarrow 0^+ }  g_p( w ) - \frac{\lambda}{p} \bigl[ (x/\lambda)^p
-1 \bigr] = 0.
\end{equation} 
Now
\ba{
\frac{\lambda}{p}
\bigl[ (x/\lambda)^p
-1 \bigr]
=
\frac{\lambda}{p}
\bigl[ \exp\bigl( p\log(x/\lambda) \bigr) -1 \bigr] 
=
\frac{\lambda}{p}
\bigl[ p\log(x/\lambda)+ o \bigl( p\log(x/\lambda) \bigr) \bigr],
}
where the little-o is as $p\log(x/\lambda)\rightarrow 0^+$, which we wish to establish as $p,\lambda\rightarrow 0^+$. Since $x > w$, we have that 
\begin{equation}
  p \log( x / \lambda ) = p \log( w / \lambda + ( x / \lambda )^{ p - 1 } ) 
   < p \log( w / \lambda + ( w / \lambda )^{ p - 1 } ) \rightarrow 0^+
\end{equation}
provided $p\rightarrow 0^+$ fast enough, such as if 
$p\sim \lambda ^q$ for any $q>0$.  This yields
\begin{multline}
  \liminf_{ ( \lambda, p ) \rightarrow ( 0^+, 0^+ ) }
\frac{g_p(2\beta)}{g_p(\alpha)}
=
\liminf_{ ( \lambda, p ) \rightarrow ( 0^+, 0^+ ) }
\frac{\lambda\log(x(2\beta)/\lambda)}{\lambda\log(x(\alpha)/\lambda)}\\
\le
\liminf_{ ( \lambda, p ) \rightarrow ( 0^+, 0^+ ) }
\frac{ \log( 2 \beta / \lambda + ( 2 \beta / \lambda )^{ p - 1 } ) }{\log(\alpha/\lambda)} 
= 
\liminf_{\lambda\rightarrow 0^+}
\frac{\log(2\beta)-\log(\lambda)}{\log(\alpha)-\log(\lambda)}
=1.
\end{multline}
Therefore, there exist $\lambda>0,p>0$ sufficiently small that $k g(2\beta)< (m+1-k) g(\alpha)$.

\paragraph{Case $p<0$} Since $g_p$ is strictly increasing on $[0,\infty)$, we take $w\rightarrow \infty$ to determine an upper bound.  Note that $x( w ) > w$ implies that $x(w)\rightarrow \infty$ as $w\rightarrow \infty$.
Then from \eqref{eq:gpexpression}, since now $p < 0$, we obtain
\ba{\label{eq:gpbound}
\lim_{w\rightarrow\infty}
g_p(w)
=
\lambda(1/2-1/p).
}
Thus for $p<0$ and all $w,\lambda$, we have $g_p(w)\le \lambda(1/2-1/p)$. Applying this with $w = 2 \beta$ and using \eqref{eq:gpexpression},
\ba{
\liminf_{\lambda\rightarrow 0^+}
\frac{g_p(2\beta)}{g_p(\alpha)}
\le&
\liminf_{\lambda\rightarrow 0^+}
\frac{\lambda(1/2-1/p)}{\lambda \bigl[ \frac1p (x(\alpha)/\lambda)^p-\frac{1}{2}(x(\alpha)/\lambda)^{2p-2}-(1/p-1/2) \bigr] }.
}
As before, $(x/\lambda)\rightarrow \infty$ as $\lambda\rightarrow 0^+$. Then
\ba{
\liminf_{\lambda\rightarrow 0^+}
\frac{g_p(2\beta)}{g_p(\alpha)}
&\le
\lim_{\lambda\rightarrow 0^+}
\frac{\lambda(1/2-1/p)}{\lambda (1/2-1/p)}=1.
}
Thus for every $p<0$ there exists $\lambda>0$ sufficiently small that $k g(2\beta)<(m+1-k)g(\alpha)$.
\end{IEEEproof}

\section{Stability}
\label{sec:stability}

Next we consider the case of noisy measurements of an approximately sparse signal.  Let $x$ be the original signal with $\|Ax-b\|_2 \le \epsilon$  whose $k$-sparse approximation is supported on $T$, i.e. $x_T=\argmin_w G(x-w) \text{ subject to } \|w\|_0=k$.  
We wish to bound $G(x^*-x)$ where
\ba{\label{eq:gineq}
x^*
=
\argmin_{w}G(w)
\text{ subject to }\|Aw-b\|_2\le \epsilon.
}
We shall bound the recovery error by the sum of a term dependent on the noise level and a term dependent on the sparse approximation error.

We shall first need two results: bounds on the magnitudes of nonzero entries of local minima of \eqref{eq:gineq} and an extension of those bounds to the error vector projected onto the null space of $A$. Recall that $\norm{ w }_{ -\infty } := \min_i \abs{ w_i }$. 

\begin{lemma}
\label{noisylb}
Assume $A\in\R^{m\times n}$ satisfies the URP and $G$ satisfies (I,II) above.  
Let $b\in \R^m$ be given. For $S \subset [ n ]$ with $|S|=m$ define $\alpha_S=\|A_S^{-1}b\|_{-\infty}$ and $\beta_S=\|A_S^{-1}b\|_\infty$.  If $\epsilon<\min_S (\alpha_S/\|A_S^{-1}\|)$, then the magnitudes of components of feasible vectors of \eqref{eq:gineq} are bounded below by $\alpha:=\min_S (\alpha_S - \|A_S^{-1}\| \epsilon ) > 0$ and bounded above by $\beta := \max_S (\beta_S+ \|A_S^{-1}\| \epsilon )$.
\end{lemma}

The assumption that $\alpha_S > 0$ for all $S$ has a similar character to the URP, in that it is true with probability 1 for random data drawn from an absolutely continuous distribution.
\begin{IEEEproof}
First, note that the error-bounded problem \eqref{eq:gineq} is equivalent to taking the $G$ minimizer from a set of equality-constrained $G$ minimizers (with different equality constraints):  For all feasible $w$, we must have $Aw=b+\eta$ for some $\|\eta\|_2\le\epsilon$.  Thus by Lemma~\ref{mnzs} the minimizer of \eqref{eq:gineq} has $m$ or fewer nonzero entries.  By the URP, any $m$ columns $S$ of $A$ give exactly one solution to $A_S w =b + \eta$.  
So we have 
\ba{
\|w\|_{-\infty} & =\|A_S^{-1}(b+\eta)\|_{-\infty}
\ge
\min_i (|A_S^{-1}b|-|A_S^{-1}\eta|)_i \nonumber \\
& \ge
\|A_S^{-1}b\|_{-\infty}-\|A_S^{-1}\eta\|_{\infty} \ge \alpha_S-\|A_S^{-1}\eta\|_2  \ge \alpha_S - \norm{ A_S^{ -1 } } \epsilon \\
& \ge \alpha, \nonumber
}
and
\ba{
\|w\|_\infty
& =
\|A_S^{-1}(b+\eta)\|_\infty
\le
\|A_S^{-1}b\|_\infty+\|A_S^{-1}\eta\|_\infty \nonumber \\
& \le \beta_S+\|A_S^{-1}\eta\|_2 \le \beta_S + \norm{ A_S^{ -1 } } \epsilon \\
& \le \beta. \nonumber
}
%% Thus the upper bound is as stated and $\epsilon < \min_S (\alpha_S/\|A_S^{-1}\|)$ guarantees the lower bound $\|y\|_{-\infty}\ge\alpha$ for the magnitude of all nonzero entries of feasible vectors.  $\alpha$ depends on $A,b,$ and sufficently small $\epsilon$.  $\beta$ depends on $A,b,$ and $\epsilon$.
\end{IEEEproof}

\begin{lemma}
\label{rprojlb}
  Assume $G$ satisfies (I,II).
  Let $x^*$ be the global minimizer of \eqref{eq:gineq},
  $x$ the original signal with $\|Ax-b\|\le \epsilon$, and let $T$ be the support of the $k$-sparse approximation of $x$. Let $\alpha_S$, $\alpha$, and $\beta$ be as in Lemma~\ref{noisylb}. Define $\alpha' := \alpha-\|x_{T^c}\|_\infty-2\epsilon$ and $\beta' := \beta+\epsilon$. 
  If $A$ satisfies the URP, $AA^T=I$,  
  $\min_S \alpha_S>\|x_{T^c}\|_\infty$ (requiring that $x$ be nearly $k$ sparse), and
  $\epsilon<\min_S \{(\alpha_S-\|x_{T^c}\|_\infty)/(2+\|A_S^{-1}\|)\}$ , 
  then the orthogonal projection $w$ of $h=x^*-x$ onto the nullspace of $A$
%% , given by $(I-A^TA)h$ (due to $AA^T=I$),
  satisfies 
  \begin{equation}
    \alpha' \le \norm{ w_{ T^c } }_{ -\infty } \text{ and } \norm{ w_{ T^c } }_{ \infty } \le 2 \beta'.
  \end{equation}
\end{lemma}

\begin{IEEEproof}
First, consider the bound 
$\epsilon<\min_S \{(\alpha_S-\|x_{T^c}\|_\infty)/(2+\|A_S^{-1}\|)\}$. 
 Note that this is stronger than the bound on $\epsilon$ from Lemma~\ref{noisylb}, and it implies $2\epsilon+\|x_{T^c}\|_\infty <\alpha$.  We see this from the following inequalities:
\ba{
\alpha
&=
\min_S \bigl\{
\alpha_S
-\epsilon\|A_S^{-1}\|
\bigr\} \nonumber \\
&>
\min_S \bigl\{
\alpha_S
-(\alpha_S-\|x_{T^c}\|_\infty)\|A_S^{-1}\|/(2+\|A_S^{-1}\|)
\bigr\} \nonumber \\
&=
\min_S \biggl\{
\frac{2\alpha_S}{2+\|A_S^{-1}\|}
+\frac{\|A_S^{-1}\|\|x_{T^c}\|_\infty}{2+\|A_S^{-1}\|}
\biggr\} \nonumber\\
&=
\min_S \biggl\{
\frac{2\alpha_S-2\|x_{T^c}\|_\infty}{2+\|A_S^{-1}\|}
+\frac{(2+\|A_S^{-1}\|)\|x_{T^c}\|_\infty}{2+\|A_S^{-1}\|}
\biggr\} \nonumber \\
&>
2\epsilon + \|x_{T^c}\|_\infty.
}
We shall use this below to guarantee $\alpha' > 0$.

Note that the hypotheses of Lemma~\ref{noisylb} are satisfied, giving $\|x^*\|_{-\infty}\ge\alpha$ and $\|x^*\|_\infty\le\beta$, $\|x\|_\infty\le\beta$. Since $A A^T = I$, the orthogonal projection of $h$ onto the nullspace of $A$ is $( I - A^T A ) h$. The desired lower bound comes from the following sequence of inequalities, using the given lower bound on nonzero elements of $x^*$, the feasibility of $x^*$ and $x$, the fact $\| A^T A \|=1$, and the assumed bound on $\epsilon$:
\bas{
\|[(I-A^TA)h]_{T^c}\|_{-\infty}
& \ge
\|h_{T^c}\|_{-\infty}
-\|A^TAh\|_\infty\\
& \ge
\|x^*_{T^c}-x_{T^c}\|_{-\infty}
-\|A^TAh\|_2\\
& \ge
\|x^*_{ T^c }\|_{-\infty}
-\|x_{T^c}\|_\infty
- \|h\|_2 \\
& \ge
\alpha
-\|x_{T^c}\|_\infty
-2 \epsilon = \alpha'>0.
}

The upper bound comes from a completely analogous argument:
\bas{
\|(I-A^TA)h\|_{\infty}
& \le
\|h\|_{\infty}
+\|A^TAh\|_\infty\\
& \le
\|x^*-x\|_{\infty}
+\|A^TAh\|_2\\
& \le
2\beta
+2\epsilon=2\beta'.
}
\end{IEEEproof}

\begin{definition}
The \emph{$G$ Noisy Nullspace Property} (or \emph{$G$ NNSP}) of order $k$ for the matrix $A$ is satisfied when for all $h\in \R^n$ and $S\subset [n]$ with $|S|\le k$ , there are constants $0\le \tau <1$ and $D\ge 0$ such that
\ba{\label{eq:nnsp}
G(h_S)
\le
\tau G(h_{S^c})
+D\|Ah\|_2
.
}
\end{definition}

\begin{proposition}
\label{GRNNSP}
Assume $G$ satisfies the triangle inequality.  
For given $A,b$, let $x^*$ be the global minimizer of \eqref{eq:gineq} and let $x$ be the original signal with $\|Ax-b\|_2\le \epsilon$  whose $k$-sparse approximation is supported on $T$.
Then the $G$ NNSP of order $k$ for $A$ implies the following stability bound:
%of $x^*$.  In particular
\ba{
G(x^*-x)
\le C_1 \epsilon + C_2 G(x_{T^c})
}
with $C_1 = 4 D / ( 1 - \tau )$ and $C_2 = 2 ( 1 + \tau ) / ( 1 - \tau )$, where $\tau$ and $D$ satisfy \eqref{eq:nnsp}.
\end{proposition}

\begin{IEEEproof}
Define the error vector  $h=x^*-x$.  Since $x^*$ and $x$ are both feasible and $\norm{ A } = 1$, $\|Ah\|_2\le 2\epsilon$.  Then by the triangle inequality of $G$,
\begin{equation}
  G(x_{T}) - G(-h_{T}) \le G(x_{T}+h_{T}).
\end{equation}
Since $G$ decouples across components,
\begin{equation}
  G(x_{T}+h_{T} ) + G( h_{T^c}) = G( x_{T}+h_{T}+h_{T^c}) = G( x^*-x_{T^c} ).
\end{equation}
Then 
\ba{
G(h_{T^c})
& \le
G(x^*-x_{T^c})
+G(h_{T})
-G(x_{T}) \nonumber \\
& \le 
G(x^*)
+G(x_{T^c})
+ G(h_{T})-G(x_{T}) \nonumber \\
& \le 
G(x)
+G(x_{T^c})+ G(h_{T})
-G(x_{T}) \nonumber \\
& =2G(x_{T^c})+G(h_{T}).\label{eq:gtrbd}
}

Now apply $G$ NNSP to $h$ on $T$:
\ba{
G(h_T)
\le 
\tau G(h_{T^c})
+D\|Ah\|_2%
\le
2\tau G(x_{T^c}) + \tau G(h_{T})
+2D\epsilon,
}
so that 
\ba{\label{eq:ht}
G(h_T)
\le
\frac{2}{1-\tau}\bigl( D\epsilon + \tau G(x_{T^c})\bigr).
}
Using \eqref{eq:gtrbd}, we obtain
\ba{\label{eq:htc}
G(h_{T^c}) \le 2G(x_{T^c}) + G(h_{T}) \le \frac{2D\epsilon}{1-\tau} + \frac{2}{1-\tau} G(x_{T^c}).
}

Now we add \eqref{eq:ht} and \eqref{eq:htc} to get the desired inequality:
\ba{
G(h)
& =
G(h_T)+G(h_{T^c}) \nonumber \\
& \le
\frac{2}{1-\tau} \bigl( D\epsilon + \tau G(x_{T^c}) \bigr) +\frac{2D}{1-\tau}\epsilon + \frac{2}{1-\tau}G(x_{T^c}) \nonumber \\
& =
\frac{4D}{1-\tau}\epsilon + \frac{2(1+\tau)}{1-\tau}G(x_{T^c}).
}
\end{IEEEproof}

\begin{theorem}[$G$ stability]
\label{Gstabrob}
Assume $A\in\R^{m\times n}$ satisfies the URP, $AA^T=I$, $G$ satisfies (I,II) above, and $G(v)\le C\sqrt{n} \|v\|_2$ for some constant $C>0$.  
For given $b$,
let $x$ be the original signal with $\|Ax-b\|_2\le \epsilon$, let $T$ be the support of its $k$-sparse approximation, and suppose  $\min_S\{\alpha_S\}> \|x_{T^c}\|_\infty$. Let $x^*$ be the global minimizer of \eqref{eq:gineq}, where  $\epsilon<\min_S \{(\alpha_S-\|x_{T^c}\|_\infty)/(2+\|A_S^{-1}\|)\}$ (with $\alpha_S$ defined as in Lemma~\ref{noisylb} ).
Define $\alpha',\beta'$ as in Lemma~\ref{rprojlb}.  
Assume that $2k < n$ and $kg(2\beta')<( n - k )g(\alpha').$
Then
\ba{
G(x^*-x)
\le
2 \biggl( 1-\frac{kg(2\beta')}{(n-k)g(\alpha')} \biggr)^{-1}
\biggl[
2C\sqrt{n}\epsilon
+\left(1+\frac{kg(2\beta')}{(n-k)g(\alpha')}\right)G(x_{T^c})
\biggr].\label{eq:gstab}
}
\end{theorem}

\begin{IEEEproof}
We shall show that the given hypotheses allow for the same application of the $G$ NNSP as in Proposition~\ref{GRNNSP}, and in a similar way, arrive at stability.  
Define $h=x^*-x$.  Since $G$ satisfies the triangle inequality, we have $G(h_{T^C})\le G(h_T) +2G(x_{T^C}),$ as in the proof of Proposition~\ref{GRNNSP}.  

Next we write $h$ as the sum of its orthogonal projections onto $\ker(A)$ and $\ker(A)^\perp$, which we denote by $w$ and $v$ respectively.  
First, suppose that there exists some $0\le \tau<1$ such that $G(w_T)\le\tau G(w_{T^c})$ (which we will prove below). Then we have: 
% Given that bound and using the definitions of $w$ and $v$, along with the fact that $A$ has all $m$ singular values equal to 1, we get the desired application of $G$ NNSP as follows:
\ba{
G(h_T)
& \le
G(w_T)+G(v_T) \nonumber \\
& \le
\tau G(w_{T^c})+G(v_T)=\tau G(w_{T^c}+v_{T^c}-v_{T^c})+G(v_T) \nonumber \\
& \le
\tau G(h_{T^c})+G(v_{T^c})+G(v_T) \nonumber \\
&=\tau G(h_{T^c})+G(v) \nonumber \\
& \le
\tau G(h_{T^C})+C\sqrt{n}\|v\|_2.\label{eq:gnnsp}
}
Since $A A^T = I$ and $v \in \ker(A)^\perp$, it follows that $v = A^T A v$. Hence $\norm{ v }_2^2 = \norm{ A v }_2^2$. Then from \eqref{eq:gnnsp} we obtain 
\ba{
G(h_T)
\le
\tau G(h_{T^c})+C\sqrt{n}\|Av\|_2.
}
And so we have the application of the $G$ NNSP to $h$ on $T$ with constants $\tau$ and $D=C\sqrt{n}.$
From here the stability inequality \eqref{eq:gstab} follows as in Proposition~\ref{GRNNSP}.

Now we go back to prove $G(w_T)\le\tau G(w_{T^c})$.  We shall use the lower bound $\|w_{T^c}\|_{-\infty}\ge\alpha'$ and the upper bound $\|w_{T}\|_\infty\le\beta'$ from Lemma~\ref{rprojlb}.  
We overestimate $G(w_T)$ and underestimate $G(w_{T^c})$ as follows:
\ba{
G(w_T)
\le 
kg(2\beta'),
\quad
G(w_{T^c})
\ge
(n-k)g(\alpha').
}
So to get $G(w_T)\le\tau G(w_{T^c})$, it suffices to have $kg(2\beta')\le\tau (n-k)g(\alpha')$, and thus $kg(2\beta')<(n-k)g(\alpha')$ guarantees some $0\le\tau<1$.  The condition $k<n-k$ gives $(n-k)/k>1$ and thus makes the inequality possible for $\alpha'<2\beta'$.

Plugging in $\tau=\frac{kg(2\beta')}{(n-k)g(\alpha')}$ to the stability inequality we get from the previous argument gives
\ba{
G(h)
\le
2 \biggl( 1-\frac{kg(2\beta')}{(n-k)g(\alpha')} \biggr)^{-1}
\biggl[
2C\sqrt{n}\epsilon
+\biggl(1+\frac{kg(2\beta')}{(n-k)g(\alpha')} \biggr) G(x_{T^c})
\biggr]
.
}
\end{IEEEproof}

\begin{corollary}[$\Gf$ stability]
\label{GFstabrob}
Assume $A\in\R^{m\times n}$ satisfies the URP, $AA^T=I$, and $G=\Gf,$ the penalty corresponding to firm thresholding.
For given $b$,
let $x$ be the original signal with $\|Ax-b\|_2\le \epsilon$  whose $k$-sparse approximation is supported on $T$, with  $\min_S\{\alpha_S\}> \|x_{T^c}\|_\infty$,
and  $x^*$ be the global minimizer of \eqref{eq:gineq}, where  $\epsilon<\min_S \{(\alpha_S-\|x_{T^c}\|_\infty)/(2+\|A_S^{-1}\|)\}$ (with $\alpha_S$ defined as in Lemma~\ref{noisylb}).
Define $\alpha',\beta'$ as in Lemma~\ref{rprojlb}.  
If $2k<n$ and $\mu < \min\{\alpha' \frac{n-k}{k}\left(1+\sqrt{1-\frac{k}{n-k}}\right),2\beta'\}$ then $x^*$ is stable, satisfying the following inequality:
\ba{
\Gf(x^*-x)
\le
2 \biggl( 1-\frac{k\gf(2\beta')}{(n-k)\gf(\alpha')}\biggr)^{-1}
\biggl[
2C\sqrt{n}\epsilon
+\biggl(1+\frac{k\gf(2\beta')}{(n-k)\gf(\alpha')}\biggr)\Gf(x_{T^c})
\biggr]
.
}
\end{corollary}

The proof of Corollary~\ref{GFstabrob} is an application of Theorem~\ref{Gstabrob} combined with the corresponding computations from the proof of Corollary~\ref{GFERP}.

\begin{corollary}[$G_p$ stability]
  \label{GPstabrob}
Assume $A\in\R^{m\times n}$ satisfies the URP, $AA^T=I$, and $G=G_p,$ the penalty corresponding $p$-shrinkage.
For given $b$,
let $x$ be the original signal with $\|Ax-b\|_2\le \epsilon$  whose $k$-sparse approximation is supported on $T$, with  $\min_S\{\alpha_S\}> \|x_{T^c}\|_\infty$,
and  $x^*$ be the global minimizer of \eqref{eq:gineq}, where  $\epsilon<\min_S \{(\alpha_S-\|x_{T^c}\|_\infty)/(2+\|A_S^{-1}\|)\}$ (with $\alpha_S$ defined as in Lemma~\ref{noisylb} ).
If $2k<n$ then there exist $0<p<1,0<\lambda$ sufficiently small so that $x^*$ is stable, satisfying the following inequality.
\ba{
G_p(x^*-x)
\le
2\biggl(1-\frac{kg_p(2\beta')}{(n-k)g_p(\alpha')}\biggr)^{-1}
\biggl[
2C\sqrt{n}\epsilon
+\biggl(1+\frac{kg_p(2\beta')}{(n-k)g_p(\alpha')}\biggr)G(x_{T^c})
\biggr]
.
}
Also, for any $p<0$ there exists $\lambda>0$ sufficiently small such that $x^*$ is stable, and the above inequality holds.
\end{corollary}

The proof of Corollary~\ref{GPstabrob} is an application of Theorem~\ref{Gstabrob} combined with the corresponding computations from the proof of Corollary~\ref{GPERP}.

\section{Convergence of iterative $p$-shrinkage}
\label{sec:convergence}

Now we consider an algorithm that employs generalized shrinkage. Consider the following optimization problem:
\begin{equation}
  \min_x F_p( x ) := \lambda G_p( x ) + \tfrac{ 1 }{ 2 } \norm{ A x - b }_2^2,
\end{equation}
where $\norm{ A } < 1$.
Applying forward-backward splitting to this problem gives \emph{iterative $p$-shrinkage} (IPS): 
\begin{equation}\label{eq:alg}
  x^{ n + 1 } = S_p( x^n - A^T ( A x^n - b ) ).
\end{equation}
This generalizes the \emph{iterative soft thresholding} algorithm (ISTA) \cite{daubechies-2004-iterative}, which is the case $p = 1$. ISTA was shown in \cite{daubechies-2004-iterative} to be globally convergent to a global minimizer (necessarily, since $F_1$ is convex). In this section, we prove global convergence of IPS for general $p < 1$, though only to a stationary point of $F_p$. Portions of the proof appeared in \cite{voronin-2013-new}, though statements there concerning convergence to a local minimizer are incorrect. 

Recall from Lemma~\ref{lem:fp} that $g_p$ is $C^\infty$ on $( 0, \infty )$. A closer examination of the proof shows that $g_p$ on $[ 0, \infty )$ is the restriction of a function that is $C^\infty$ on $\R$, so $g_p$ is one-sided differentiable to all orders at $w = 0$.

The following follows exactly as in the known case of $p = 1$ \cite{daubechies-2004-iterative}:
\begin{lemma}[\cite{voronin-2013-new}]\label{lem:decrease}
  Let $\lambda > 0$ and $p \in \R$, and define $\{ x^n \}$ by \eqref{eq:alg}, with $x^0$ arbitrary. 
  \begin{enumerate}
    \item $F( x^{ n + 1 } ) \le F( x^n )$ for all $n$, and $F( x^{ n + 1 } ) < F( x^n )$ unless $x^n$ is a fixed point of the algorithm.
    \item $\norm{ x^{ n + 1 } - x^n }_2 \rightarrow 0$.
  \end{enumerate}
\end{lemma}

\begin{lemma}
  Let $\lambda > 0$ and $p \in \R$. The fixed points of \eqref{eq:alg} are precisely the stationary points of $F_p$.
\end{lemma}
\begin{IEEEproof} 
  The iteration \eqref{eq:alg} can be seen as minimizing the surrogate functional 
  \begin{equation}
    \lambda G_p( x ) + \tfrac{ 1 }{ 2 } \norm{ A x - b }_2^2 + \tfrac12 \norm{ x - w }_2^2 - \tfrac12 \norm{ A x - A w }_2^2
  \end{equation}
  with fixed $w = x^n$, by expanding the quadratic terms and rearranging to express the minimizer in terms of the proximal mapping of $G_p$. Therefore the first-order optimality condition of this functional is satisfied at $x = x^{ n + 1 }$. Also, the first-order optimality condition of this functional at $x = x^n$
is the same as the first-order optimality condition of $F_p$ at $x = x^n$. Hence $x^{ n + 1 } = x^n$ if and only if the first-order optimality condition of $F_p$ at $x = x^n$ is satisfied.
\end{IEEEproof}

The lemma shows why it is not possible to show that IPS converges to a local minimizer: if the algorithm happens to be initialized with a stationary point that is not a local minimizer (\emph{i.e.}, a saddle point or local maximizer), then the initializer is a fixed point of the algorithm, so the algorithm cannot converge to a local minimizer in such a case.

\begin{lemma}
  Fix $\lambda > 0$, $p \in ( -\infty, 1 )$. We have $g_p''' > 0$ on $( 0, \infty )$, $g_p''' < 0$ on $( -\infty, 0 )$, $g_p'''( 0+ ) > 0$, and $g_p'''( 0- ) < 0$.
\end{lemma}
\begin{IEEEproof}
  Since $g_p$ is even, it suffices to consider $w > 0$.
  Above we had that $x = x( w ) =  ( f_p^* )'( w )$ satisfies $x - \lambda^{ 2 - p } x^{ p - 1 } = w$. Differentiating with respect to $w$, we have that 
  \begin{equation}\label{eq:tprime}
    x' - \lambda^{ 2 - p } ( p - 1 ) x^{ p - 2 } x' = 1,
  \end{equation} 
  so 
  \begin{equation}\label{eq:xprime}
    x' = \bigl( 1 - \lambda^{ 2 - p } ( p - 1 ) x^{ p - 2 } \bigr)^{ -1 }.
  \end{equation}
  Since $p < 1$, $( f_p^* )''( w ) = x'( w ) > 0$ for all $w > 0$.

  Differentiating \eqref{eq:tprime}, we get
  \begin{equation}
    x'' - \lambda^{ 2 - p } ( p - 1 ) \bigl[ ( p - 2 ) x^{ p - 3 } ( x' )^2 + x^{ p - 2 } x'' \bigr] = 0,
  \end{equation}
  or
  \begin{equation}\label{eq:xprimeprime}
    x'' \bigl( 1 - \lambda^{ 2 - p } ( p - 1 ) x^{ p - 2 } \bigr) = \lambda^{ 2 - p } ( p - 1 ) ( p - 2 ) x^{ p - 3 } ( x' )^2,
  \end{equation}
  implying that $x''$ has the same sign as $x$. Since $x( w )$ has the same sign as $w$, we have that $( f_p^* )'''( w )$ has the same sign as $w$ for $w \ne 0$.

  Differentiating the relation \eqref{eq:gp} defining $g_p$, we obtain $w + \lambda g_p'( w ) = ( f_p^* )'( w )$, $1 + \lambda g_p''( w ) = ( f_p^* )''( w )$, and $\lambda g_p'''( w ) = ( f_p^* )'''( w )$. Thus $g_p'''( w )$ has the same sign as $w$ for $w \ne 0$ as well. Also, $\lambda g_p'''( 0+ ) = ( f_p^* )'''( 0+ ) = \lim_{ w \rightarrow 0^+ } x''( w )$. Since $\lim_{ w \rightarrow 0^+ } x( w ) = \lambda$, we obtain from \eqref{eq:xprime} and \eqref{eq:xprimeprime} that $( f_p^* )'''( 0+ ) = \frac{ 1 - p }{ ( 2 - p )^2 } \lambda^{ -1 } > 0$. Thus $g'''( 0+ ) > 0$.
\end{IEEEproof}

\begin{lemma}
  Let $p \ge 0$. Then $\{ x^n \}$ is bounded.
\end{lemma}
\begin{IEEEproof}
  Since $\{ F_p( x^n ) \}$ decreases monotonically, it suffices to show that $F_p$ is coercive, which we establish be showing coercivity of $g_p$. By \eqref{eq:xwsp}, if $w\rightarrow \infty$, then $x \rightarrow \infty$. For $p > 0$, that $g_p( w ) \rightarrow \infty$ follows from \eqref{eq:gpexpression}. The $p = 0$ case is similar, but $f_0$ has a different form:
\ba{
g_0( w ) 
&= 
( x w - f_0( x ) - w^2 / 2 ) / \lambda \nonumber \\
&= 
\bigl( x w - x^2 / 2 + \lambda^{ 2 } \log x - \lambda^2 ( \log \lambda - 1 / 2 ) - w^2 / 2 \bigr) / \lambda \nonumber \\
&=
 \lambda \log x - ( x - w )^2 / ( 2 \lambda ) - \lambda ( \log \lambda - 1 / 2 ) \nonumber \\
&=
\lambda \log x  -\tfrac{\lambda}{2} (x/\lambda)^{-2}- \lambda ( \log \lambda - 1 / 2 ). 
}
From this the coercivity of $g_0$ follows.
\end{IEEEproof}

\begin{lemma}\label{lem:bound}
  Let $p < 0$, and assume $\lambda^2 > p \norm{ b }_2^2 / ( p - 2 )$. Let $x^0 = 0$. Then $\{ x^n \}$ is bounded.
\end{lemma}
\begin{IEEEproof}
  From Lemma~\ref{lem:decrease}, we know that $F_p( x^n )$ decreases (strictly except at a fixed point, in which case we are done). Then for $n \ge 1$,
  \begin{equation}
    F_p( x^n ) < F_p( x^0 ) = \norm{ b }^2_2 / 2,
  \end{equation}
  so 
  \begin{equation}\label{eq:gbound}
    G_p( x^n ) \le F_p( x^n ) / \lambda < \norm{ b }^2_2 / ( 2 \lambda ).
  \end{equation}
  By \eqref{eq:gpbound}, $g_p( w ) < ( 1 / 2 - 1 / p ) \lambda$. Combining this bound with \eqref{eq:gbound}, we obtain for each $j$,
  \begin{equation}
    g_p( x^n_j ) \le G_p( x^n ) < \norm{ b }^2_2 / ( 2 \lambda ) < ( 1 / 2 - 1 / p ) \lambda.
  \end{equation}
  Letting $t$ be the unique positive number satisfying $g( t ) = \norm{ b }^2_2 / ( 2 \lambda )$, we obtain $\norm{ x^n }_\infty < t$ independently of $n$.
\end{IEEEproof}

Now we can establish convergence of our algorithm.
\begin{theorem}
Let $\lambda > 0$, $p \in ( -\infty, 1 )$. Let the sequence $\{ x^n \}$ be defined by \eqref{eq:alg}, with $x^0$ arbitrary for $p \ge 0$, and $x^0 = 0$ for $p < 0$ in which case we further assume $\lambda^2 > p \norm{ b }_2^2 / ( p - 2 )$. Then $\{ x^n \}$ converges to a stationary point of $F$.
\end{theorem}

\begin{IEEEproof}
  We have that $F_p( x^{ n + 1 } ) < F_p( x^n )$ unless $x^n$ is a fixed point, $F$ is continuous, and the sequence $\{ x^n \}$ is bounded. Then by \cite[Thm.~3.1]{meyer1976sufficient}, we have that either $\{ x^n \}$ converges or its limit points form a continuum. (A continuum is a compact, connected set; here we also exclude the degenerate case of a singleton.) Since we already know that any limit point of $\{ x^n \}$ will be a stationary point of $F_p$, we complete the proof by showing that the stationary points of $F_p$ cannot form a continuum.

  Let $E$ be the set of stationary points of $F_p$, and suppose $E$ is a continuum. Fix $\bar{x} \in E$. For any $\epsilon > 0$, it cannot be that $\N( \bar{x}; \epsilon ) \cap E = \{ \bar{x} \}$, otherwise $\{ \bar{x} \}$ would be both open and closed in $E$, contrary to $E$ being connected. Thus there is a sequence of stationary points $\bar{x} + v^n$ with $v^n \ne 0$, $v^n \rightarrow 0$.  

  Since $\{ v^n / \norm{ v^n } \}$ is a sequence of unit vectors, it cannot converge to zero. Then we can fix $j$ such that $\{ v^n_j / \norm{ v^n } \}$ does not tend to zero, though of course $v^n_j \rightarrow 0$. First suppose that $\bar{ x }_j \ne 0$. By considering a tail of $v^n_j$, we can assume that $\bar{x}_j + v^n_j \ne 0$ for all $n$. Then $g_p$ is differentiable at $\bar{x}_j$ and $\bar{x}_j + v^n_j$, and since $\bar{x}$ and $\bar{x} + v^n$ are fixed points, 
\begin{equation}
  \lambda^{ 2 - p } g_p'( \bar{x}_j + v^n_j ) + \bigl[ A^T( A ( \bar{x} + v^n ) - b ) \bigr]_j = 0
\end{equation} and 
\begin{equation}
  \lambda^{ 2 - p } g_p'( \bar{x}_j ) + \bigl[ A^T( A \bar{x} - b ) \bigr]_j = 0.
\end{equation} 

  Define $\fee( x ) = \lambda g_p'( x_j ) + \bigl[ A^T( A x - b ) \bigr]_j$. All derivatives of $\fee$ exist at every $x \ne 0$. Letting $(a_i)$ denote the columns of $A$, if $i \ne j$, we have $\partial \fee /\partial x_i ( \bar{x} ) = \langle a_i, a_j \rangle$, while $\partial \fee / \partial x_j ( \bar{x} ) = \lambda g''( \bar{x}_j ) + \norm{ a_j }^2$. Also, $\fee( \bar{x} ) = 0$ and each $\fee( \bar{x} + v^n ) = 0$. By differentiability of $\fee$, we have 
  \begin{equation}\label{eq:difffee}
    \frac{ \fee( \bar{x} + v^n ) - \fee( \bar{x} ) - \nabla \fee( \bar{x} ) \cdot v^n }{ \norm{ v^n } } \rightarrow 0.
  \end{equation}  
  Since the first two terms of \eqref{eq:difffee} are zero, $\nabla \fee( \bar{x} ) \cdot v^n  = o( \norm{ v^n } )$ as well. By continuity of $\nabla \fee$ at $\bar{x}$, it is straightforward to show that $\nabla \fee( \bar{x} + v^n ) \cdot v^n = o( \norm{ v^n } )$ also.

  Now we consider second derivatives. $\partial^2 \fee / \partial x_i \partial x_k ( \bar{x} ) = 0$, unless $i = k = j$, while $\partial^2 \fee / \partial x_j^2 ( \bar{x} ) = \lambda g_p'''( \bar{x}_j )$. Now by the differentiability of $\nabla \fee$, 
  \begin{equation}
    \norm{ \nabla \fee( \bar{x} + v^n ) - \nabla \fee( \bar{x} ) - \nabla^2 \fee( \bar{x} ) \ v^n } = o( \norm{ v^n } ),
  \end{equation}
  so 
  \begin{equation}
    \nabla \fee( \bar{x} + v^n ) \cdot v^n - \nabla \fee( \bar{x} ) \cdot v^n - v^n \cdot \nabla^2 \fee( \bar{x} ) \ v^n = o( \norm{ v^n }^2 ).
  \end{equation} 
  But from the above we have that the first two terms are $o( \norm{ v^n }^2 )$, so $v^n \cdot \nabla^2 \fee( \bar{x} ) \ v^n = o( \norm{ v^n }^2 )$ as well. But this is $\lambda g_p'''( \bar{x}_j ) ( v^n_j )^2$; since $( v^n_j )^2 / \norm{ v^n }^2$ does not tend to zero by choice of $j$, it must be that $g_p'''( \bar{x}_j ) = 0$, a contradiction. 

  Thus we must have $\bar{x}_j = 0$. By choice of $j$, infinitely many $v^n_j \ne 0$, so by passing to a subsequence we may assume that either all $v^n_j > 0$ or $v^n_j < 0$. By the one-sided differentiability of $g_p$, we can then repeat the above argument using a smooth extension of $g_p$ to $\R$. Since neither $g_p'''(0+)$ nor $g_p'''(0-)$ are zero, we will obtain the same contradiction. Therefore $E$ cannot be a continuum, and the sequence $\{ x^n \}$ defined by \eqref{eq:alg} is  convergent to a stationary point of $F_p$.
\end{IEEEproof}

\section{Conclusion}
\label{sec:conclusion}

We have shown that for given signals with reasonable sparsity assumptions and a broad class of measurement matrices, the families of penalties corresponding to $p$-shrinkage and firm thresholding, like the $\ell^p$ quasinorms, provide a candidate penalty that is able to exactly recover the given data with the given measurement matrix.  Further we have shown that these penalties behave well with respect to the addition of noise in the measurements, or only approximately sparse signals (as is often the case in practical settings).  Finally, we have shown that iterative $p$-shrinkage converges to stationary points of the unconstrained energy.  These results, together with empirical results (see \cite{chartrand-2014-shrinkage}, and Fig.~\ref{fig:SL}), further support the idea that generalized shrinkage penalties can be an advantageous alternative to standard $\ell^1$ compressed sensing, or $\ell^p$ compressed sensing.  

Further work could benefit from exploring in what generality these type of results hold.  The theory of generalized shrinkage allows for an endless possibility of other shrinkages and penalties to study.  Additionally, the methods of proof may apply to compressed sensing relaxations that arise in other ways.  Generally speaking, determining conditions under which convex optimization results can be extended to handle nonconvex functionals may continue to be a fruitful area of research.  Lastly, we make no claims that the approximations made in these proofs give the tightest results possible, so further refinement of these results may be possible and interesting.

% or


\appendices

\ifCLASSOPTIONcompsoc
  \section*{Acknowledgments}
\else
  \section*{Acknowledgment}
\fi

The authors would like to acknowledge the support of the UC Lab Fees Research grant 12-LR-236660 in conducting this research.  
The first author also acknowledges the support of NSF grant no. DGE-1144087, and would like to thank his graduate advisor, Professor Andrea L. Bertozzi, and his other LANL mentor, Brendt Wohlberg, for their guidance.
The second author also acknowledges the support of the U.S.\ Department of Energy through the LANL/LDRD Program.

\ifCLASSOPTIONcaptionsoff
  \newpage
\fi



%\bibliographystyle{IEEEtran}
\bibliographystyle{IEEEtran}
\bibliography{bib}


\begin{IEEEbiography}[{\includegraphics[width=1in,height=1.25in,clip,keepaspectratio]{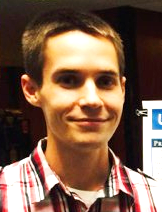}}]{Joseph Woodworth}
received a B.Sc.(Hons.) degree in mathematics from the University of Maryland, College Park in 2011.  He is a fourth year graduate student at the Department of Mathematics at the University of California, Los Angeles in pursuit of a Ph.D. under the supervision of Professor Andrea L. Bertozzi.  His research interests include variational methods, nonlocal operators, image processing,  density estimation,  networks, dictionary learning, compressed sensing, and optimization.
\end{IEEEbiography}

\begin{IEEEbiography}[{\includegraphics[width=1in,height=1.25in,clip,keepaspectratio]{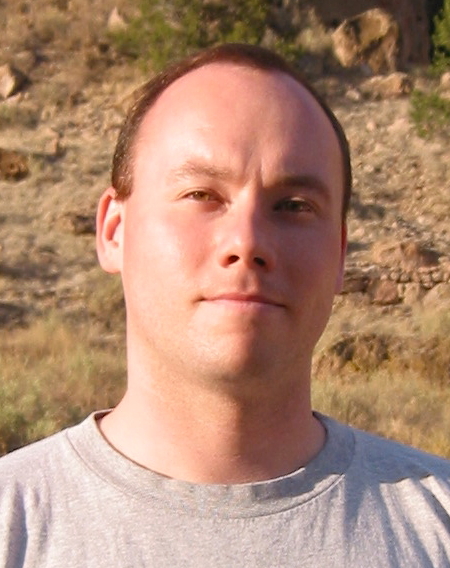}}]{Rick Chartrand}
(M'06--SM'12) received a B.Sc.(Hons.) degree in mathematics from the University of Manitoba in 1993, and a Ph.D. in mathematics from the University of California, Berkeley in 1999. He held academic positions at Middlebury College and the University of Illinois at Chicago before coming to Los Alamos National Laboratory in 2003, where he is currently a technical staff member in the Applied Mathematics and Plasma Physics group. His research interests include compressive sensing, nonconvex continuous optimization, image processing, dictionary learning, computing on accelerated platforms, and geometric modeling of high-dimensional data.
\end{IEEEbiography}

%\newpage


%\vfill


\end{document}